\documentclass[%
 reprint,
amsmath,amssymb,
 aps,
]{revtex4-2}

\usepackage{graphicx}% Include figure files
\usepackage{dcolumn}% Align table columns on decimal point
\usepackage{bm}% bold math
\usepackage{amssymb}
\usepackage{amsmath}
\usepackage{mathptmx}
\usepackage{etoolbox}
\usepackage[dvipsnames]{xcolor}
\usepackage{gensymb}
\begin{document}

\preprint{APS/123-QED}

\title{The Narrow Escape Problem of a Chiral Active Particle (CAP): An Optimal Scheme}% Force line breaks with \\
\thanks{Email: sathish.akella@iitjammu.ac.in}%

\author{Alakesh Upadhyaya}
 \altaffiliation[Also at ]{Department of Physics,Indian Institute of Technology Jammu, Jammu and Kashmir, India,181221.}%Lines break automatically or can be forced with \\
\author{Venkata Sathish Akella}%

\affiliation{
 Department of Physics,Indian Institute of Technology Jammu, Jammu and Kashmir, India,181221
}%

\begin{abstract}
We report a simulation study on the narrow escape kinetics of a Chiral Active Particle (CAP) confined to a circular domain with a narrow escape opening. The study's main objective is to optimize the CAP's escape chances as a function of the relevant parameters, such as translational and rotational speeds of the CAP, domain size, etc. We identified three regimes in the escape kinetics \emph{namely} the noise-dominated regime, the optimal regime, and the chiral activity-dominated regime. In particular, the optimal regime is characterized by an escape scheme that involves a direct passage to the domain boundary at first and then a unidirectional drift along the boundary towards the exit. Furthermore, we propose a non-dimensionalization approach to optimize the escape performance across microorganisms with varying motile characteristics. Additionally, we explore the influence of the translational and rotational noise on the CAP's escape kinetics.

\end{abstract}

\maketitle

%\tableofcontents

%%%MAIN TEXT%%%%
\section{Introduction}
Active matter systems can extract energy from their environment, exhibiting spatiotemporal structures well beyond equilibrium states~~\cite{Bechinger-RMP-2016}. This deviation from equilibrium gives rise to an array of novel phenomena that distinguish them from their counterparts in thermal equilibrium. Noteworthy among these are collective phenomena like flocking and swarming, among others~\cite{Marchetti-RMP-2013, Bar-ARCMP-2020, Zhou-CommPhys-2022}. The exploration of active systems spans diverse length and time scales, encompassing entities like flocks of birds~\cite{Cavagna-ARCMP-2014}, biological micro-swimmers~\cite{Tung-SR-2017}, granular particles~\cite{Kumar-NC-2014}, chemo-tactic Janus particles~\cite{Vutukuri-NC-2020}, and paramagnetic nanoparticles ~\cite{Yu-NC-2018}. Such investigations hold the potential to uncover a plethora of new properties, thereby providing a foundational basis for the advancement of synthetic intelligent materials and devices. One salient feature of the active systems is their ability to manifest distinct behaviors contingent upon the geometric attributes of their confinement. Henceforth, the study of active systems confined within enclosed spaces has garnered substantial attention from various groups since the past decades~\cite{Wioldand-PRL-2013,Caprini-JCP-2021,Lushi-PNAS-2014, Thakur-PRE-2012,Ortiz-PRL-2005,Fily-SM-2014,Takatori-NC-2016, Negi-PRR-2023, Thery-SciRep-2020,Aranson-PRL-2022}.

The studies above have demonstrated the emergence of distinct collective dynamics due to the confinement effect. Despite these investigations' valuable insights, numerous aspects of active systems remain unexplored, prompting the need for further research. One such scenario arises when confined particles possess a finite probability of exiting a closed domain. This phenomenon holds significant importance in fields like biology and biophysics, exemplified by the diffusion of ions in cellular micro-domains~\cite{Schuss-PNAS-2007, Holcman-SIAM-2014}, and is commonly referred to as \textit{``The Narrow Escape Problem''}. The narrow escape problem endeavors to study and understand the escape kinetics of these particles through a small escape aperture in their respective confinement, and these kinetics are typically characterized by mean first passage times to the target.

Narrow escape problems for active systems have been extensively investigated theoretically, encompassing both two-dimensional and three-dimensional domains~\cite{Cheviakov-PRE-2012,Srivastava-PRE-2021, Singer-PRE-2008, Paoluzzi-PRE-2020, Singer-JSP-2006,Gomez-PRE-2015,Tzou-MMS-2015,Ridgway-PRE-2019, Zhang-Entropy-2023,Grebenkov-PCCP-2016, Guerin-PRE-2023, Benichou-PRL-2008, Mangeat-PRE-2021}. Recent simulations have highlighted the utility of narrow escape problems in segregating active particles with distinct motility characteristics ~\cite{Paoluzzi-PRE-2020}. Notably, simulations involving Viscek particles~\cite{Olsen-PRR-2020}with interactions have revealed that the mean first passage time exhibits a non-exponential decay under low noise conditions. Conversely, in noise-dominated scenarios, the escape time displays exponential decay~\cite{Olsen-PRR-2020}, resembling the behavior of non-interacting particles. Further, a simulation study on confined active particles with single and multiple exit apertures~\cite{Debnath-JCP-2021, Nayak-JCP-2020} explored the escape kinetics in underdamped and overdamped regimes under different noise conditions. Likewise, in a separate simulation investigation~\cite{Kumar-SM-2023}, active nanorods confined within a circular region with single and multiple escape openings highlighted the critical significance of the nanorod's activity and its persistence length in optimizing their escape time. Simultaneously, experimental endeavors involving confined undulating worms~\cite{Biswas-SM-2023} placed in interconnected circular chambers via a narrow channel escape a chamber by consistently traversing along the boundary of the chamber. So far in the above literature and to the best of our knowledge the effect of chiral activity on narrow escape kinetics has not been explored. We believe such a study will offer valuable insights into both experimental and theoretical studies for further exploration. 

Towards this, we computationally investigate the escape kinetics of Chiral Active Particles (CAPs) confined to a two-dimensional circular chamber featuring a narrow escape opening. Our goal is to find optimal combinations of parameters involving rotational speed (strength of chiral activity), translational speed, and domain size such that the escape time for the CAP is minimal. We generalize our findings using a non-dimensionlization approach involving the parameters mentioned earlier. Further, we study the effect of noise on the escape kinetics and briefly compare the findings to Active Brownian Particles. \par
The structure of this paper unfolds as follows: In Section \ref{Model}, we delve into the specifics of our employed model. Moving on to Section~\ref{sec:results}, we present our findings, and in Section~\ref{sec:conclusion}, we provide concluding remarks along with suggestions for potential directions for future research.

\section{Simulation Details}
\label{Model}
\begin{figure}[h!]
\centering
\includegraphics[width=0.9\linewidth]{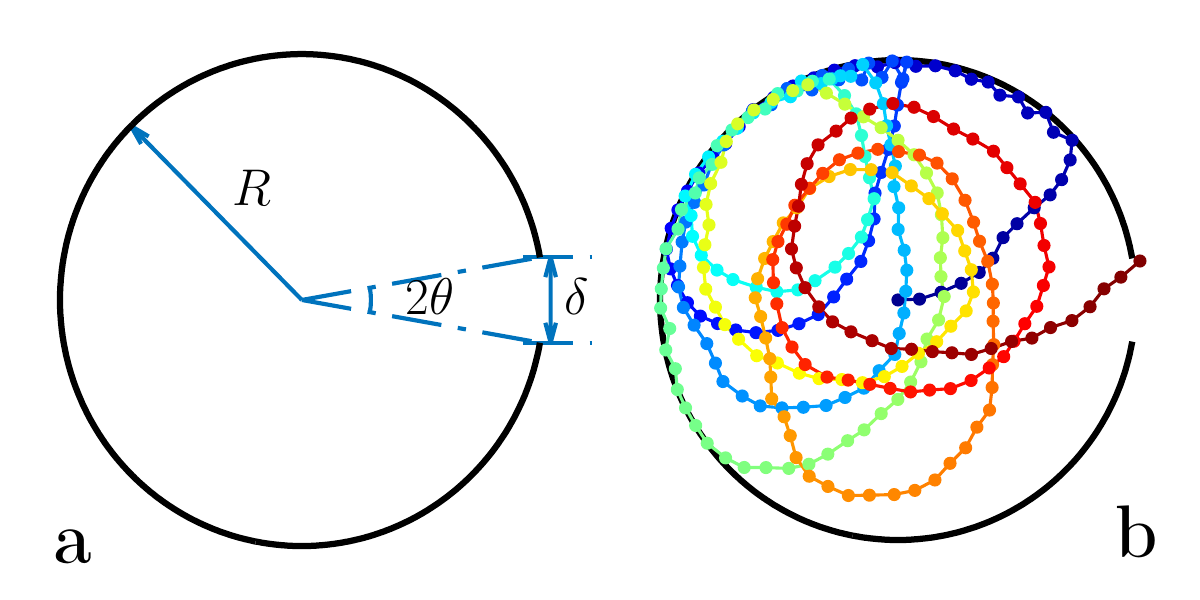}
\caption{(a) A circular confinement of radius $R$ with an exit angle $2\theta$ and the corresponding exit width $\delta$. (b) Typical escape trajectory of a Chiral Active Particle. Color scheme blue to red indicates the passage of time.}
\label{fig:figure1}
\end{figure}

We study the escape kinetics of a microscopic and spherical chiral active particle (CAP) confined to a circular chamber with a narrow escape opening in two dimensions. Fig.~\ref{fig:figure1} shows the schematic of the narrow escape problem. We confine a chiral active particle to a circular chamber of radius $R$ with an exit angle $2\theta$ (which corresponds to an exit width $\delta$). The study aims to optimize the escape performance as a function of system parameters, such as translational speed, rotational speed, domain size, and thermal noise. Typically, the Reynolds number associated with the motion of microorganisms in an aqueous environment is of the $O(10^{-4})$~\cite{Purcell-AJP-1977}, i.e., the inertial forces are negligible compared to viscous forces. Therefore, we employ Brownian dynamics to simulate the system. We mathematically model and simulate the dynamics of the CAP as described in the referred works.~\cite{Volpe-AJP-2014,Liebchen-EPL-2022}. In brief, the CAP's equations of motion are described by overdamped Langevin's equations and are:

\begin{subequations}\label{eq:eom}
\begin{align} 
\dot{\vec{r}} &= v \hat{\theta} + \sqrt{2\mathrm{D}_{T}}\vec{\xi}(t) \label{eq:eomt} \\
\dot{\theta} &= \omega + \sqrt{2\mathrm{D}_{R}}\zeta(t) \label{eq:eomr}
\end{align}  
\end{subequations} 

Where $\vec{r}(t)$ and $\theta(t)$ are the instantaneous position and orientation of the particle, $\hat{\theta} = (\cos{\theta(t)}, \sin{\theta(t)})$ is the unit vector in the direction of instantaneous orientation. Also, $\mathrm{D}_{T}$ and $\mathrm{D}_{R}$ are the diffusion constants characterizing translational and rotational diffusion processes, respectively, of the CAP. The translational speed $v$ and rotational speed $\omega$ characterize the motility of a CAP. Further, $\vec{\xi} = (\xi_{x}, \xi_{y})$ and $\zeta$ are Gaussian white noise variables with zero mean and unit variance. Following Volpe et al.~\cite{Volpe-AJP-2014}, we chose $\mathrm{D}_{T} = 0.22\ \mu\mathrm{m^2/s}$ and $\mathrm{D}_{R} = 0.16\ \mathrm{rad^2/s}$ for a spherical particle of $1\ \mu\mathrm{m}$ in size, unless otherwise mentioned. These values correspond to a typical bacterium (like \textit{E. Coli}) swimming in an aqueous medium at $T = 300\ \mathrm{K}$. In addition, we defined translational and rotational P\'{e}clet numbers as $\mathrm{Pe}_{T} = v/\sqrt{\mathrm{D}_{T}\mathrm{D}_{R}}$ and $\mathrm{Pe}_{R} = \omega/\mathrm{D}_{R}$ respectively to characterize the relative strength of deterministic versus stochastic contributions to the motility of the CAP. For all the cases reported in this study, $\mathrm{Pe}_{T} > 70$ unless otherwise mentioned. Also, the sign of chiral activity $\omega$, i.e., clockwise or counterclockwise rotation, has no bearing on our results. In the following sections, we discussed the escape kinetics as we varied $v$, $\omega$, and $R$ over typical ranges corresponding to a real system~\cite{Bechinger-RMP-2016}.

To simulate the CAP's motion, we integrate the equations of motion (Eq.~\eqref{eq:eom}) using the Euler-Maruyama scheme with a time step $\mathrm{d}t = 0.01\ \mathrm{s}$. We start the simulation at the center of the domain with initial orientation $\hat{\theta}_{0} = (\cos{\theta_{0}}, \sin{\theta_{0}})$, where $\theta_{0}$ is randomly picked from a uniform distribution over $[0, 2\pi]$. When the particle encounters the boundary, we implement reflective boundary conditions (as described in~\cite{Volpe-AJP-2014}). A CAP's typical escape trajectory with sufficiently high chiral activity (i.e., $\omega$) is shown in Fig.~\ref{fig:figure1}b (color scheme blue to red indicates passage of time). We characterize an escape event using $\tau$ as the time the CAP takes to escape the domain (also known as \textit{mean first passage time}). In general, $\tau$ is a function of translational speed $v$, rotational speed $\omega$, domain size $R$, exit angle $\theta$ and the diffusion coefficients $\mathrm{D}_{T}$, $\mathrm{D}_{R}$. We fixed $\theta$ = $10\degree$ and varied other relevant parameters in all our simulations. Since the exit angle is fixed at $\theta = 10\degree$, the exit width $\delta$ proportionately varies with the domain size $R$; however, we observed that this variation has little effect on our findings for the domain sizes considered in this study. Further, to obtain a statistical estimation of escape times, we simulated five trials, with $N = 1000$ escape events each, for a given set of parameters characterizing the CAP's motion. We define the mean escape time $\bar{\tau}$ for a trial as:

\begin{equation}
\mathrm{\bar{\tau}= \frac{1}{N}\sum_{i=1}^{N} \tau_{i} }
\end{equation}  

Where, $\tau_{i}$ is the escape time for the $i^{\mathrm{th}}$ event. In the following sections, we discuss and conclude the escape kinetics as relevant parameters are varied.

\begin{figure*}[ht!]
\centering
\includegraphics[width=0.95\linewidth]{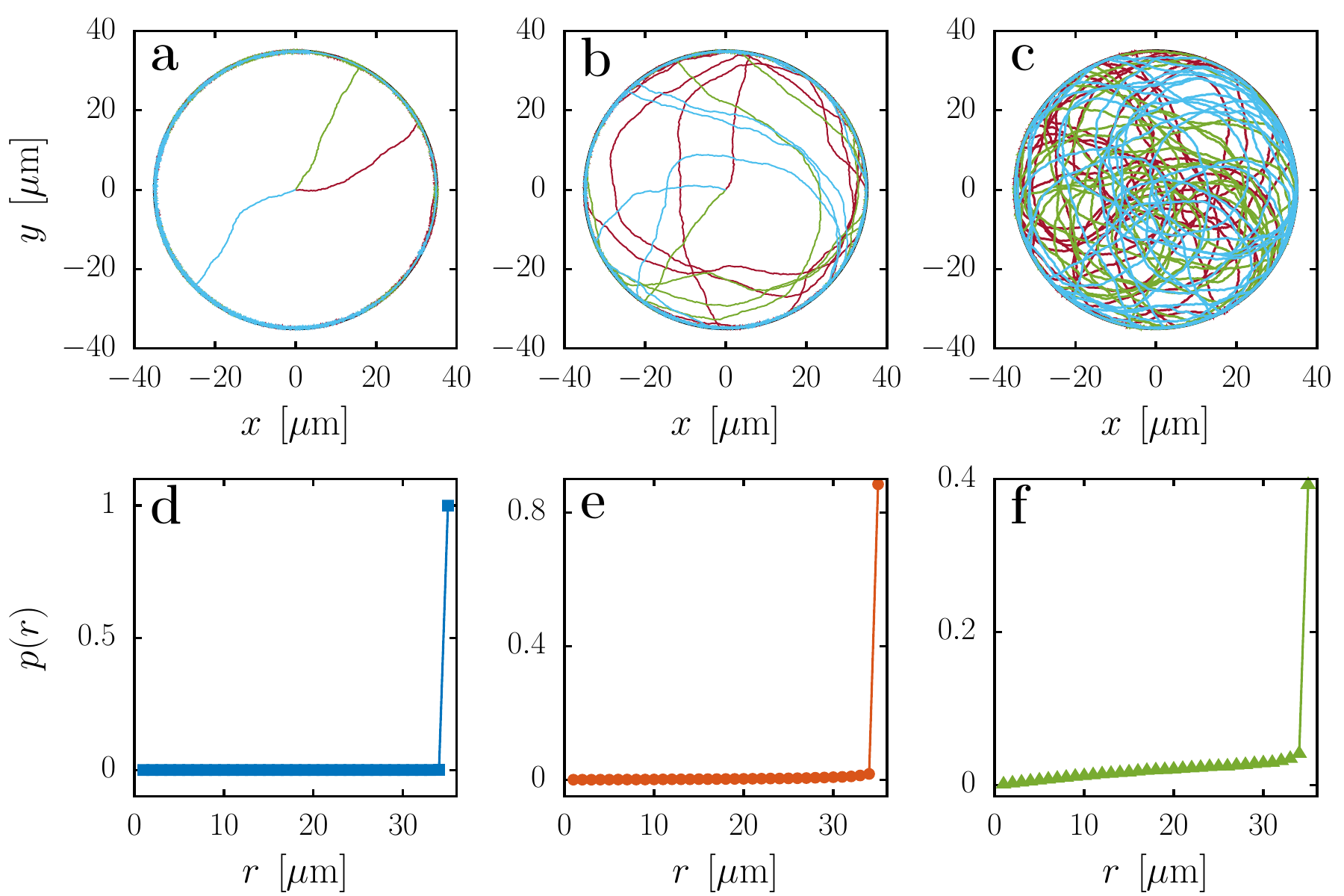}
\caption{Motion trajectories and the corresponding averaged probability distribution functions as function of radial distance of a CAP starting from three different initial orientations at (a) \& (d) $\omega = 0\ \mathrm{rad/s}$ (b) \& (e) $\omega = 0.628\ \mathrm{rad/s}$ and (c) \& (f) $\omega = 1.256\ \mathrm{rad/s}$. In all the three cases the translational speed and the domain size are fixed at $v = 31.4\ \mu\mathrm{m/s}$ and $R=35\ \mu\mathrm{m}$ respectively.} 
\label{fig:figure2}
\end{figure*}

\section{Results and Discussion}
\label{sec:results}

\begin{figure*}[ht]
\centering
\includegraphics[width=0.9\linewidth]{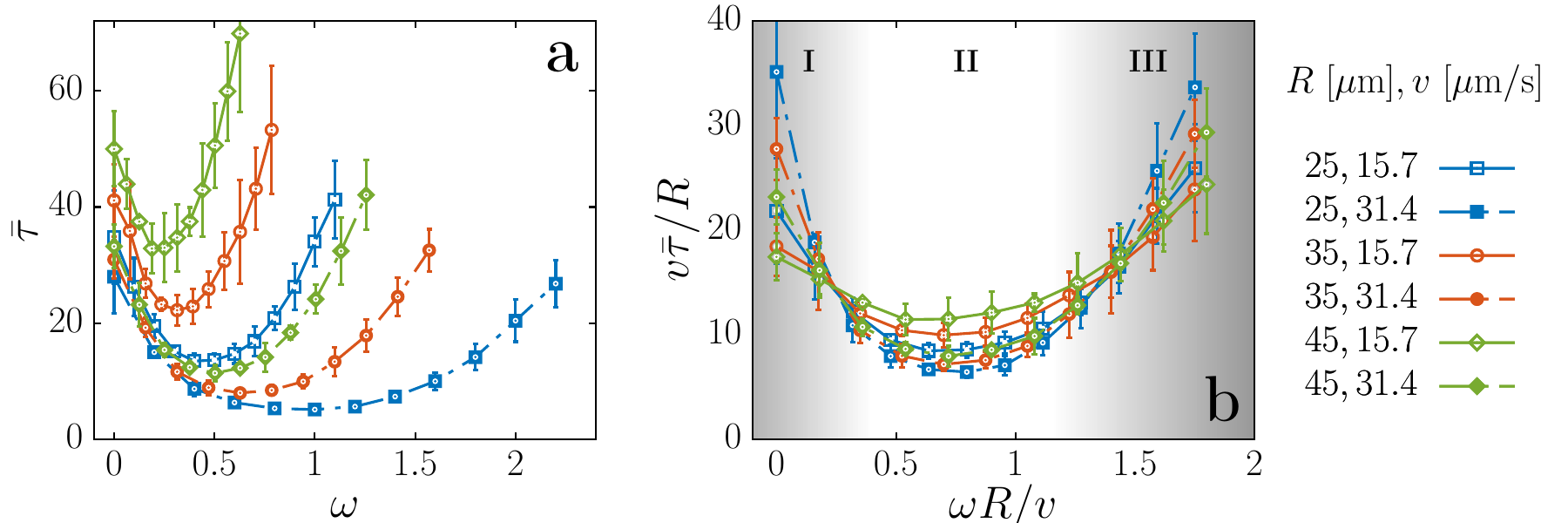}
\caption{(a) Mean escape times vs. chiral activity $\omega$. (b) Non-dimensionalized mean escape times vs. non-dimensionalized parameter $\omega R/v$. The error bars, in both the figures, are plotted at 10x the actual error for clarity.}
\label{fig:figure3}
\end{figure*}

Active Brownian Particles (ABPs) are a mathematical model to simulate achiral microorganisms and their motion is described by Eq.~\eqref{eq:eom} with $\omega = 0$. ABPs correctly reproduce the ``\textit{wall-hugging}'' nature~\cite{Bricard-NC-2015} that is commonly observed in these microorganisms. As expected, when the chiral activity is small in comparison to thermal noise (i.e., when $\mathrm{Pe}_{R} < 1$), the CAPs also exhibit the wall-hugging behavior, similar to ABPs. However, when $\mathrm{Pe}_{R} \gg 1$, the chiral activity is large enough to lift off the particle from the boundary. Fig.~\ref{fig:figure2}a-c show typical escape trajectories exhibited by the CAPs at $\omega = 0\ \mathrm{rad/s}$, $\omega = 0.628\ \mathrm{rad/s}$ and $\omega = 1.256\ \mathrm{rad/s}$ respectively for a constant domain size $R = 35\ \mu\mathrm{m}$ and translational speed $v = 31.4\ \mu\mathrm{m/s}$. These trajectories show that the particle's lift-off occurrences increase with increasing chiral activity $\omega$. Further, this behavior is also reflected in the probability distribution function (PDF) of the particle's radial location from the center of the domain. Fig.~\ref{fig:figure2}d-f show respective averaged PDFs corresponding to the trajectories in Fig.~\ref{fig:figure2}a-c to confirm the same.

From the above discussion, we deduce that there are mainly two schemes by which a CAP typically escapes the domain, excluding those that directly reach the exit. First, the CAP directly reaches the domain boundary and traverses \emph{mostly} along the domain boundary until the escape. In addition, the CAP's passage along the boundary is generally a quasi-one-dimensional process because it mostly strays close to the boundary. Second, the CAP explores the domain landscape more often and occasionally may have short passages along the boundary but escapes mostly because of random drifts toward the exit. For example, the scenarios presented in Fig.~\ref{fig:figure2}a(d),b(e) correspond to the former scheme, whereas Fig.~\ref{fig:figure2}c(f) corresponds to the latter.

In order to study and understand these escape kinetics in detail, we simulated the narrow escape problem for a chiral active particle as a function of relevant parameters. Fig.~\ref{fig:figure3}a shows the mean escape times $\bar{\tau}$ as a function of chiral activity $\omega$ for different domain sizes $R$ and different translational speeds $v$ at $\mathrm{D}_{T} = 0.22\ \mu\mathrm{m^2/s}$ and $\mathrm{D}_{R} = 0.16\ \mathrm{rad^2/s}$. As evident from the figure, there is a minimum in all the simulated cases. One noteworthy observation is that the particle exhibits tighter curvature trajectories as $\omega$ increases for a constant $v$ (see the supplementary movie named \textit{regime-III.avi}). A measure of the radius of curvature of a CAP's trajectory is $R_{\mathrm{CAP}} = v/\omega$. Though apparent, it is worthwhile to note that the radius of curvature of the trajectories decreases with increasing chiral activity; therefore, one would expect the escape times to be large at high chiral activity as the particle spends substantial time circling inside the domain before exiting. Our observations also qualitatively agree with this understanding. This enabled us to define a non-dimensional parameter $R/R_{\mathrm{CAP}} = \omega R/v$ to study systems ranging across different length scales. For example, $\omega R/v = 2$ corresponds to a scenario where the domain size is twice the radius of curvature of the particle's trajectory. Towards this, we replotted the mean escape times in Fig.~\ref{fig:figure3}a as a function of the non-dimensional parameter $\omega R/v$ and is shown in Fig.~\ref{fig:figure3}b. Note that the mean escape times are normalized by $R/v$ to account for their dependence on domain size, $R$, where $R/v$ is the time taken by the CAP if it were to travel a distance $R$ straight towards the exit with a translational speed $v$ and escape the domain.

\subsection{Escape Kinetics}
The non-dimensionalization of the chiral activity and mean escape times yields a universal feature in escape kinetics across various conditions. We especially note (see Fig.~\ref{fig:figure3}b) that there is an optimal value of $\omega R/v \sim 0.8$ for which the escape time is minimal across all the cases. In other words, the minimum corresponds to $R_\mathrm{CAP} \sim 1.25R$, i.e., the radius of curvature of the CAP's trajectory is approximately 25\% larger than the domain size. A careful observation of the trajectories corresponding to $\omega R/v \sim 0.8$ reveals that the particle follows the most optimal escape scheme (other than a direct path towards the exit) and has two characteristic features: (1) a direct passage to the domain boundary and (2) a steady and unidirectional drift towards the exit along the boundary (see the supplementary movie named \textit{regime-II.avi}). The sign of chiral activity determines the direction of the drift and has \emph{no} effect on escape times. Typically, this is also the escape scheme for cases of $\omega R/v \lesssim 0.4$ except that the particle mostly \emph{wanders} rather than steadily drifting along the boundary in one direction (see the supplementary movie named \textit{regime-I.avi}). The escape scheme for the cases when $\omega R/v \gtrsim 1.2$ is a combination of multiple factors, including drifting along the boundary, accidental passages towards the exit, etc. (see the supplementary movie named \textit{regime-III.avi}). We divided the kinetics into three regimes \textit{viz.} I. noise-dominated regime II. the optimal regime, and III. chiral activity-dominated regime and are discussed in detail in the following sections.

\subsubsection{I. Noise-dominated Regime}
\label{sec:regimea}
The noise-dominated regime is highlighted and labeled as (I) in Fig.~\ref{fig:figure3}b. This regime is observed when $\omega R/v \lesssim 0.4$. As mentioned, this regime is characterized by a direct passage to the domain boundary, and then the particle wanders along the boundary till escape. In this regime, the particle spends significant time wandering along the boundary. The time taken by the CAP to reach the boundary (\emph{via} the directed motion) is \emph{only} a small fraction of the mean escape time. Thus, wandering motion along the boundary is the limiting process. Note that translational diffusion (\emph{via} $D_\mathrm{T}$) plays an insignificant role as the translational P\'{e}clet number $\mathrm{Pe}_{T} \gg 1 $ in all the cases reported in Fig.~\ref{fig:figure3} however that is not the case with rotational diffusion. When the CAP encounters the domain boundary, it wanders until its orientation (which is determined by Eq.~\eqref{eq:eomr}) is directed away from the boundary. Since $\mathrm{Pe}_{R} \lesssim 1$, the CAP mostly stays ``hugged'' to the boundary (similar to an ABP) and eventually escapes the domain.

\subsubsection{II. Optimal Regime}
\label{sec:regimeb}
This regime is characterized by an optimal scheme for the particle's escape from the domain. It involves two observable features: (1) a direct passage to the domain boundary and (2) a steady and unidirectional drift along the boundary. For a given domain size $R$ and translational speed $v$, there is an optimal value for the chiral activity w.r.to the rotational noise such that the particle continues to drift unidirectionally along the boundary till the escape and hardly lifts off the boundary during the whole process. This regime is highlighted and labeled as (II) in Fig.~\ref{fig:figure3}b and occurs for $0.4 \lesssim \omega R/v \lesssim 1.2$

\subsubsection{III. Chiral Activity-dominated Regime}
\label{sec:regimec}
In contrast to regimes (I) and (II), this regime is characterized by dominant chiral activity, marked by a higher value of $\omega$ that results in higher escape times from the domain. As mentioned earlier, in this regime, the radius of curvature of the CAP's trajectory is smaller than the domain size, i.e., $R_\mathrm{CAP} < R$. As a result the particle mostly explores most of the domain and is occasionally observed in contact with the boundary (for example, see Fig.~\ref{fig:figure1}b). The particle's escape is a combination of short drifts along the boundary and/or accidental excursions towards the exit. We observed that this regime occurs when $\omega R/v \gtrsim 1.2$ and is highlighted and labeled as (III) in Fig.~\ref{fig:figure3}b.

\subsection{Role of Noise}
\label{sec:noiserole}
\begin{figure}[hb!]
\centering
\includegraphics[width=0.9\linewidth]{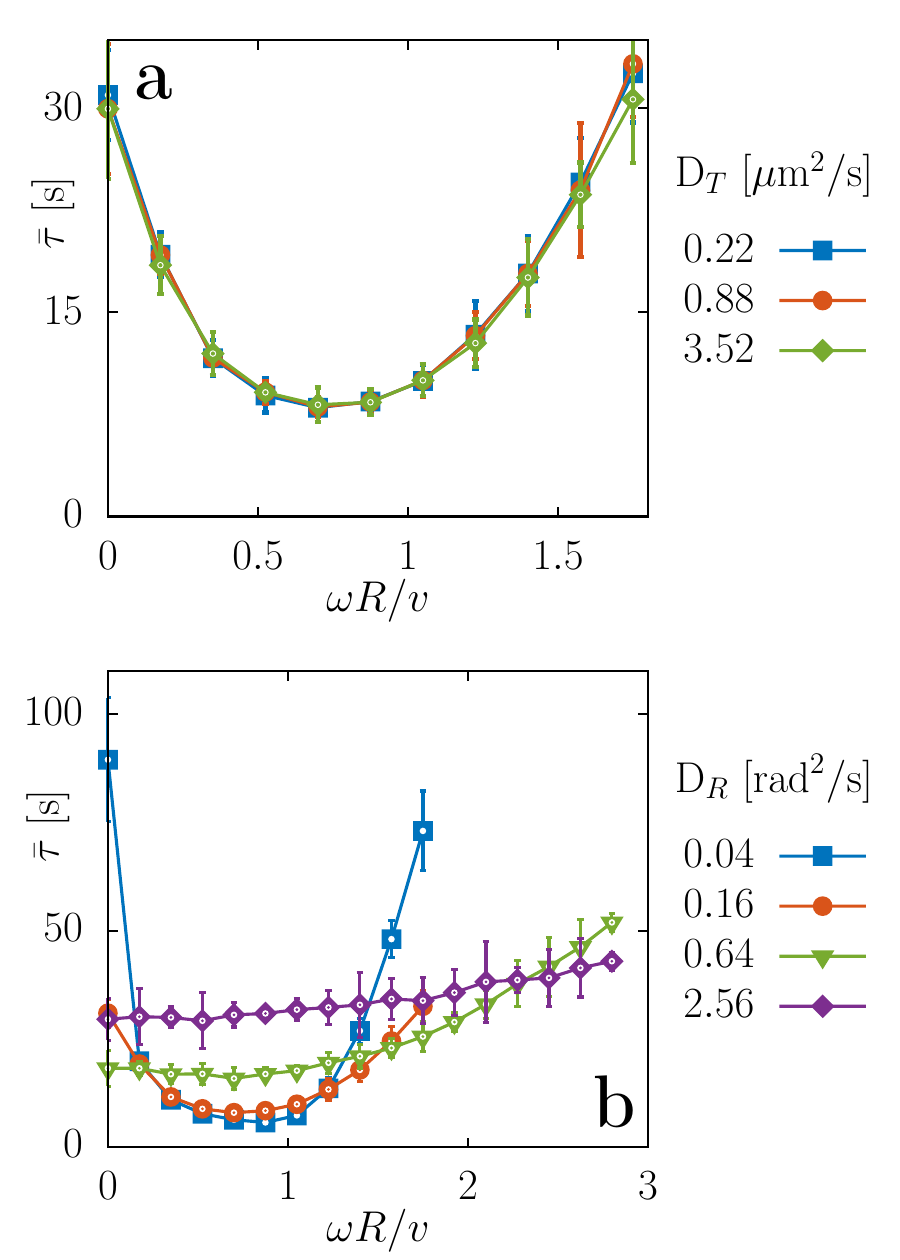}
\caption{Variation of mean escape times (a) $\bar{\tau}$ vs. $D_\mathrm{T}$ at $D_\mathrm{R} = 0.16\ \mathrm{rad^2/s}$ and (b) $\bar{\tau}$ vs. $D_\mathrm{R}$ at $D_\mathrm{T} = 0.22\ \mathrm{\mu m^2/s}$. Simulation parameters for both the cases $v = 31.4\ \mathrm{\mu m/s}$, $R = 35\ \mathrm{\mu m}$. The error bars, in both the figures, are plotted at 10x the actual error for clarity.}
\label{fig:figure4}
\end{figure}

\begin{figure}[h!]
\centering
\includegraphics[width=0.9\linewidth]{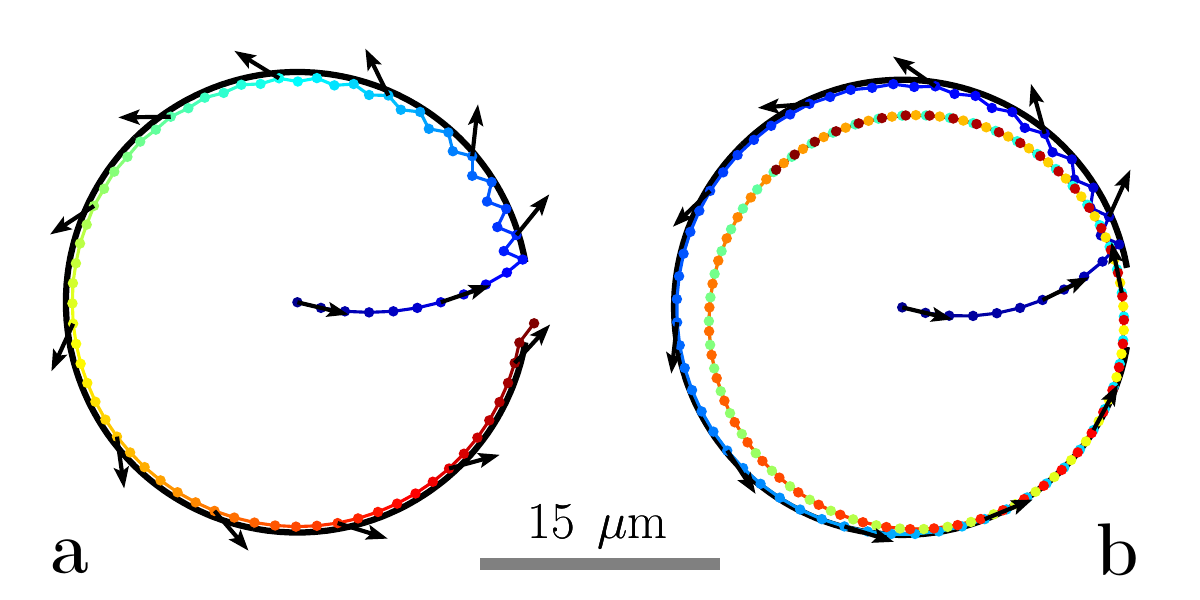}
\caption{A CAP's ideal trajectories when $\mathrm{D}_{T} = \mathrm{D}_{R} = 0$ at (a) $\omega R/v = 0.9$ and (b) $\omega R/v = 1.1$. In both the figures arrows represent the CAP's instantaneous orientation as it traverses along the boundary. Simulation parameters: (a) $\omega \sim 1.9\ \mathrm{rad/s}$, $v = 31.4\ \mu\mathrm{m/s}$ and $R = 15\ \mu\mathrm{m}$ and (b) $\omega \sim 2.3\ \mathrm{rad/s}$, $v = 31.4\ \mu\mathrm{m/s}$ and $R = 15\ \mu\mathrm{m}$.}
\label{fig:figure5}
\end{figure}

The thermal noise (\emph{via} $\mathrm{D}_{T}$ and $\mathrm{D}_{R}$ in Eq.~\eqref{eq:eom}) plays a significant role in escape kinetics. For example, achiral particles (i.e., particles with $\omega = 0$) when $\mathrm{D}_{T} = \mathrm{D}_{R} = 0$ never escape the domain, unless the CAP is oriented directly towards the exit, as they keep pushing against the boundary indefinitely. We observed that the rotational diffusion coefficient $\mathrm{D}_{R}$ has a more predominant effect on escape kinetics than the translational diffusion coefficient $\mathrm{D}_{T}$. As mentioned earlier, $\mathrm{Pe}_{T} \gg 1$ for all the cases reported in this study, therefore translational noise is expected to have minimal effect on escape times. Moreover, in regimes I and II, the CAP meanders close to the boundary during most of its escape course and keeps pushing against the boundary until it gets oriented (determined by Eq.~\eqref{eq:eomr}) away from the boundary. Once at the boundary, there is hardly any effect of translational diffusion. In contrast, the rotational component governed by the second part of the overdamped Langevin's equations (Eq.~\eqref{eq:eomr} decides whether the CAP wanders/drifts along unidirectionally or lifts off the boundary. Now, to study the effect of these noises separately, we performed simulations with varying either $\mathrm{D}_{T}$ or $\mathrm{D}_{R}$ while the other is held constant. Fig.~\ref{fig:figure4}a shows the effect of translational diffusion coefficient $\mathrm{D}_{T}$ on mean escape times while $\mathrm{D}_{R} = 0.16\ \mathrm{rad^{2}/s}$ for a fixed translational speed $v = 31.4\ \mathrm{\mu m/s}$ and fixed domain size $R = 35\ \mathrm{\mu m}$. The figure shows that there is hardly any effect on mean escape times though $\mathrm{D}_{T}$ is varied by 16 times. On the other hand, Fig.~\ref{fig:figure4}b shows the mean escape times as $\mathrm{D}_{R}$ is varied while $\mathrm{D}_{T}$ is fixed at $0.22\ \mathrm{\mu m^{2}/s}$ for the same translational speed $v$ and domain size $R$. As can be seen from the figure, the effect of $\mathrm{D}_{R}$ is significant. Firstly, consider the cases when the particle has no chiral activity, i.e., $\omega = 0$ (and hence $\omega R/v = 0$). These cases correspond to an ABP, and we conclude that large noise levels are unfavorable, but moderate noise increases the chances of escape, in agreement with reported literature [Ref]. Secondly, at low or moderate noise, CAPs outperform ABPs at escaping the domain for optimally tuned chiral activity. The corresponding escape kinetics follow the optimal regime discussed in section ~\ref{sec:regimeb}. Nevertheless, at large chiral activities, ABPs perform better as the escape times are shorter at $\omega R/v = 0$ than at $\omega R/v > 2$. The optimal regime completely disappears after increasing the noise level further, and the mean escape times monotonically increase. Thus, CAPs perform as good or worse than ABPs, depending on the chiral activity. Further, the disappearance of the optimal regime can be explained as follows. Initially, the noise is large enough to dominate the chiral activity and completely destroys the unidirectional drift along the boundary observed at lower noise strengths. However, as the chiral activity begins to take over, the kinetics are already in regime III, i.e., the radius of curvature of the CAP's motion is comparable to or less than the domain size, and the escape scenario worsens beyond, resulting in monotonically increasing escape times. A further detail worth mentioning is that the minimum at $\sim 0.8$ asymptotically approaches $\omega R/v = 1$ as noise levels are reduced further (see the blue squares in Fig.~\ref{fig:figure4}b). To understand this behavior, consider two ideal cases with \emph{no} thermal noise (i.e. $\mathrm{D}_{T} = \mathrm{D}_{R} = 0$) at two different values of $\omega R/v = 0.9$ and $1.1$. In these scenarios, the CAP's motion is completely deterministic, and in both cases, the particle's orientation $\hat{\theta}$ directed outwards the domain upon first collision with the boundary. With time, the orientation evolves at a rate $\omega$ and begins to turn inwards. In the former case, i.e., when $\omega R/v = 0.9$, the CAP \emph{never} completely turns inwards, causing it to escape when it arrives at the exit (see Fig.~\ref{fig:figure5}a). Whereas in the latter case, i.e., when $\omega R/v = 1.1$, the CAP completely turned inwards, causing a lift-off from the boundary at a grazing angle. From then onwards, the particle's trajectory is forever confined (see Fig.~\ref{fig:figure5}b) as the radius of curvature of the CAP's motion trajectory is less than the domain radius. Additionally, the initial collision site along the boundary also affects an escape event as a CAP colliding closer to the exit may still be oriented outwards the domain, causing a swift escape. However, when averaged over a large number of initial orientations $\hat{\theta}_{0}$, collisions with the initial site far away from the exit dominate the overall escape events resulting in large escape times. The rationale provided above for the ideal cases also applies in the presence of thermal noise; however, multiple collision and lift-off events along the boundary are smeared out by the noise, resulting in a minimum in escape times at $\omega R/v < 1$ and the escape times exhibit a sharp rise beyond $\omega R/v = 1$.

\section{Conclusion}
\label{sec:conclusion}
We simulated the narrow escape kinetics of a Chiral Active Particle (CAP) confined to a circular domain with a narrow escape aperture in two dimensions using Brownian Dynamics. We showed that there exists an optimal combination of parameters such as translational speed $v$, rotational speed $\omega$, and domain size $R$ such that the escape time is minimal. We defined a non-dimensional parameter $\omega R/v$ to study and characterize the escape kinetics across different length scales. We established that the CAP's escape chances are the highest when $\omega R/v \sim 0.8$. Further, we classified these dynamics into three regimes \textit{viz} (I). noise-dominated when $\omega R/v \lesssim 0.4$ (II). optimal when $0.4 \lesssim \omega R/v \lesssim 1.2$ and (III). chiral activity-dominated when $\omega R/v \gtrsim 1.2$. Different escape schemes with characteristic features characterize these regimes. Especially in regime II, the CAP escapes the domain by a direct passage to the boundary, followed by a unidirectional drift along the boundary. In the chiral activity-dominated regime, the CAP spends much time spiraling (with the trajectory's radius of curvature less than the domain size) inside the domain before exiting accidentally, thus resulting in large escape times. Furthermore, we studied the effect of noise (via translational and rotational diffusion coefficients) and showed that rotational diffusion has a predominant effect over the escape kinetics. In particular, the optimal regime completely disappears at large noise strengths as the unidirectional motion along the boundary is subdued. Finally, we remark that these results can be experimentally verified by studying the escape kinetics of magnetotactic bacteria (in response to an externally applied magnetic field) or genetically mutated microorganisms (to achieve variable chiral activity) in microfluidic chambers with narrow escape apertures. Further, we believe these findings provide valuable insight into segregating a mixture of microorganisms with different chiral activities by appropriately choosing a domain size.
 
\section*{Author Contributions}
Conceptualization by AU and VSA; Data Curation, Formal Analysis, Investigation, Methodology, Supervision, Validation and Visualization by VSA; Writing - Original Draft by AU, Writing - Review \& Editing by VSA.

\section*{Conflicts of interest}
There are no conflicts to declare.

\section*{Acknowledgements}
AU would like to thank Ministry of Human Resources and Development (MHRD), Govt. of India for financial assistance. VSA acknowledges the financial support (Seed grant SGT-100037 ) from IIT Jammu.
%%%END OF MAIN TEXT%%%
\providecommand*{\mcitethebibliography}{\thebibliography}
\csname @ifundefined\endcsname{endmcitethebibliography}
{\let\endmcitethebibliography\endthebibliography}{}

\end{document}